\documentclass[final,5p,times,twocolumn]{elsarticle}

\usepackage{lineno}
\usepackage[hidelinks]{hyperref}

\usepackage{graphicx}
\usepackage{amsmath}   
\usepackage[section]{placeins}
\begin{document}

\begin{frontmatter}

%% Note: \pmbanner before the actual title
\title{Design of an 8-Channel 40 GS/s 20 mW/Ch Waveform Sampling ASIC in 65 nm CMOS}

\author[1]{Jinseo Park\corref{cor1}
  %\fnref{fn1}
  }
\ead{truewis@uchicago.edu}

\author[7]{Evan Angelico}

\author[1]{Andrew Arzac}

\author[2]{Davide Braga}
\author[1]{Ahan Datta}
\author[2]{Troy England}
\author[4]{Camden Ertley}
\author[2]{Farah Fahim}
\author[1]{Henry J. Frisch}
\author[1]{Mary Heintz}
\author[3]{Eric Oberla}
\author[2]{Nathaniel J. Pastika}
\author[6]{Hector D. Rico-Aniles}
\author[2]{Paul M. Rubinov}
\author[2]{Xiaoran Wang}
\author[1]{Yui Man Richmond Yeung}
\author[2]{Tom N. Zimmerman}
 \cortext[cor1]{Corresponding author}
 %\fntext[fn1]{This is the first author footnote.}

\affiliation[1]{organization={Enrico Fermi Institute, the University of Chicago}, 
                 addressline={933 East 56th Street},
                 postcode={60637}, 
                 city={Chicago, IL}, 
                 country={USA}}
\affiliation[2]{organization={Fermi National Accelerator Laboratory},
                 postcode={60510}, 
                 city={Batavia, IL}, 
                 country={USA}}
\affiliation[3]{organization={Kavli Institute for Cosmological Physics, the University of Chicago}, 
                 addressline={5640 South Ellis Avenue},
                 postcode={60637}, 
                 city={Chicago, IL}, 
                 country={USA}}
\affiliation[4]{organization={Southwest Research Institute}, 
                 addressline={6220 Culebra Road},
                 postcode={78238}, 
                 city={San Antonio, TX}, 
                 country={USA}}

\affiliation[6]{organization={North Central College}, 
                 addressline={30 N. Brainard Street},
                 postcode={60540}, 
                 city={Naperville, IL}, 
                 country={USA}}

\affiliation[7]{organization={Stanford University}, 
                 addressline={450 Jane Stanford Way},
                 postcode={94305}, 
                 city={Stanford, CA}, 
                 country={USA}}

\begin{abstract}
1 ps timing resolution is the entry point to signature based searches relying on secondary/tertiary vertices and particle identification. We describe a preliminary design for PSEC5, an 8-channel 40 GS/s waveform-sampling ASIC in the TSMC 65 nm process targetting 1 ps resolution at 20 mW power per channel.  
Each channel consists of four fast and one slow switched capacitor arrays (SCA), allowing ps time resolution combined with a long effective buffer. Each fast SCA is 1.6 ns long and has a nominal sampling rate of 40 GS/s. The slow SCA is 204.8 ns long and samples at 5 GS/s. Recording of the analog data for each channel is triggered by a fast discriminator capable of multiple triggering
during the window of the slow SCA.
To achieve a large dynamic range, low leakage, and high bandwidth, the SCA sampling switches are implemented as 2.5 V nMOSFETs controlled by 1.2 V shift registers. Stored analog data are digitized by an external ADC at 10 bits or better.

Specifications on operational parameters include a 4 GHz analog bandwidth and a dead time of 20 microseconds,
 corresponding to a 50 kHz readout rate, determined by the choice of the external ADC.
\end{abstract}

\begin{keyword}
Waveform-sampling, ADC, Picosecond, ASIC, 4 GHz bandwidth, 65nm CMOS
\end{keyword}

\end{frontmatter}

%text of the article

%% Use \section commands to start a section
\section{Introduction}
\FloatBarrier

1 ps timing resolution is the entry point to signature based searches relying on secondary / tertiary vertices and particle identification. In addition, multiple hit capability and a long time buffer are desirable. An essential requirement for large fast electronics systems is a low power consumption per channel.

With a bandwidth of 4 GHz and a sampling rate of 40 GS/s, the maximum time resolution of an incoming pulse is predicted to be better than 1 ps. The architecture of the chip is designed to provide a high sampling rate with a long buffer as well as multi-hit capability.

The preliminary design of PSEC5 is in TSMC 65 nm process. Simulations predict a maximal power consumption during sampling to be roughly 20 mW/Ch. Each channel consists of four fast and one slow switched capacitor arrays (SCA). This initial version of the chip uses external ADCs. The external ADCs determine the readout rate; there are 1280 sampling capacitors per channel, read out serially once per event. Using a 40MHz ADC per channel, this gives roughly 32 $\mu$s of the dead time per event. 

\begin{table}[htb]%% placement specifier
%% Use tabular environment to tag the tabular data.
%% https://en.wikibooks.org/wiki/LaTeX/Tables#The_tabular_environment
\centering%% For centre alignment of tabular.
\begin{tabular}{|c|c|}%% Table column specifiers
%% Tabular cells are separated by &
\hline
   Process & 65 nm TSMC \\ %% A tabular row ends with \\
\hline
  Signal to Noise Ratio & 1000 \\
\hline
  Sampling Rate & 40 GS/s(5 GS/s) \\\hline
  Buffer Length & 6.4 ns(204.8 ns) \\\hline
  Analog Bandwidth & 4 GHz \\\hline
  Channels & 8 \\\hline
  Area & 2.4 $\text{mm}^2$ \\\hline
\end{tabular}
%% Use \caption command for table caption and label.
\caption{PSEC5 Specifications.}\label{spec}
\end{table}
\begin{figure*}[htb]
\centering
\includegraphics[width=\linewidth]{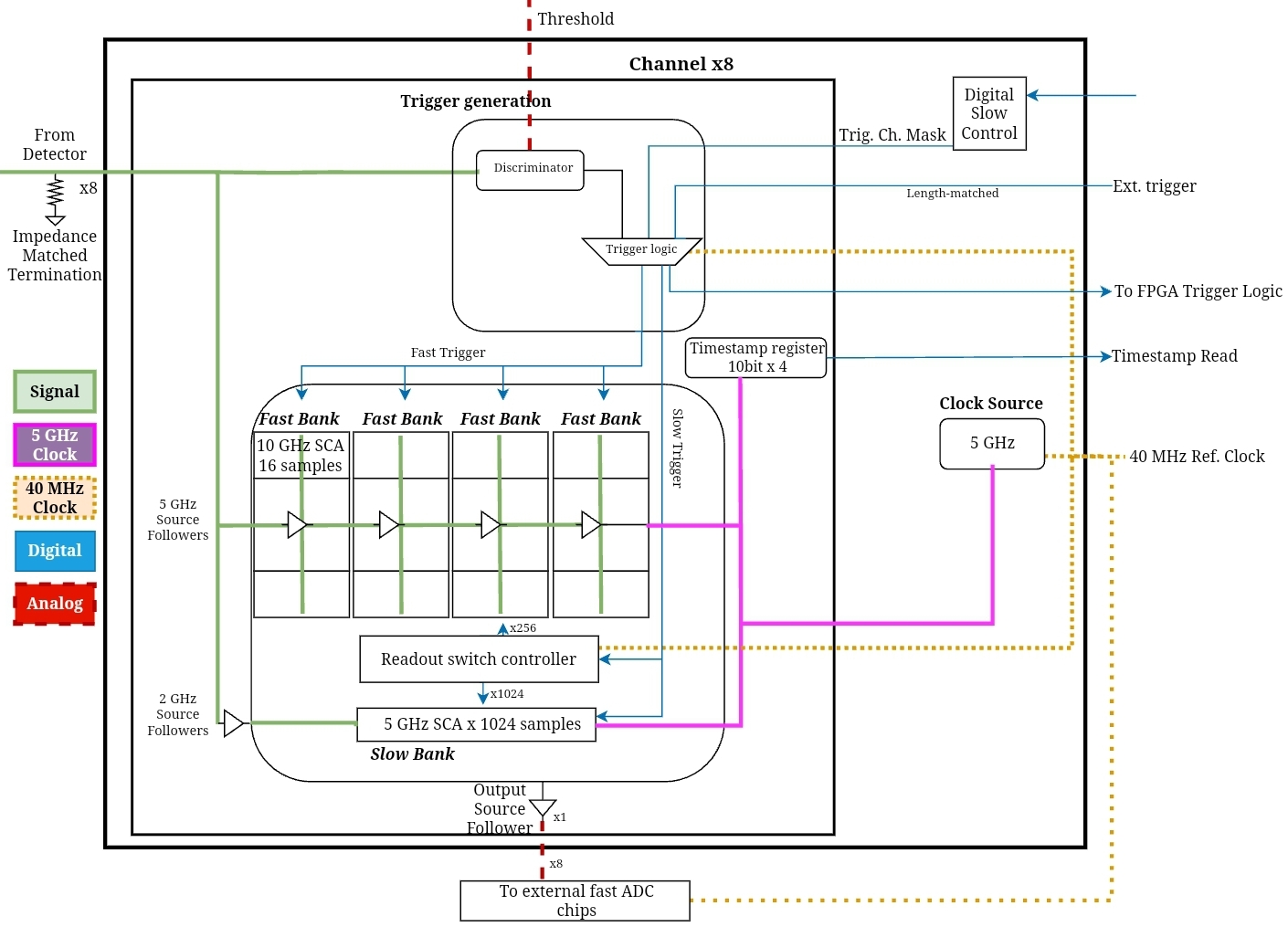}
\caption{Functional Block Diagram.}\label{psec5Overview}
\end{figure*}
The organization of the paper is as follows.
We present the functional block diagram in section \ref{BlockDiagramSection}, and fast and slow switched capacitor array (SCA) columns in section \ref{SCASection}. We present the source followers we put to achieve higher analog bandwidth in section \ref{signalSection}. The layout details are elaborated in section \ref{layoutSection}. Power consumption and source follower simulation results are shown in section \ref{simSection}.

\section{Block Diagram}
\label{BlockDiagramSection}
The block diagram is shown in Fig \ref{psec5Overview}. Each channel consists of a discriminator that generates fast triggers on the rising edges of the signal.

\section{Switched Capacitor Array}
\label{SCASection}
Switched capacitor arrays (SCA) consist of sampling capacitors which sequentially sample the voltage of the signal line.
\subsection{Fast and Slow SCA}
There are four Fast SCAs per channel, 64 samples each. This provides four times 1.6 ns of the sampling window, which can be configured to either run sequentially to capture multiple rising edges of the signal or run as a single longer SCA. The slow SCA is 1024 samples long (204.8 ns) and starts sampling before the fast SCAs. The fast buffers' triggered positions within the slow bank are timestamped. This architecture prevents cumulative timing errors while achieving a long sampling window.

\subsection{Voltage Level Shifter}

\FloatBarrier

\begin{figure}[htb]
\centering
\includegraphics[width=\linewidth]{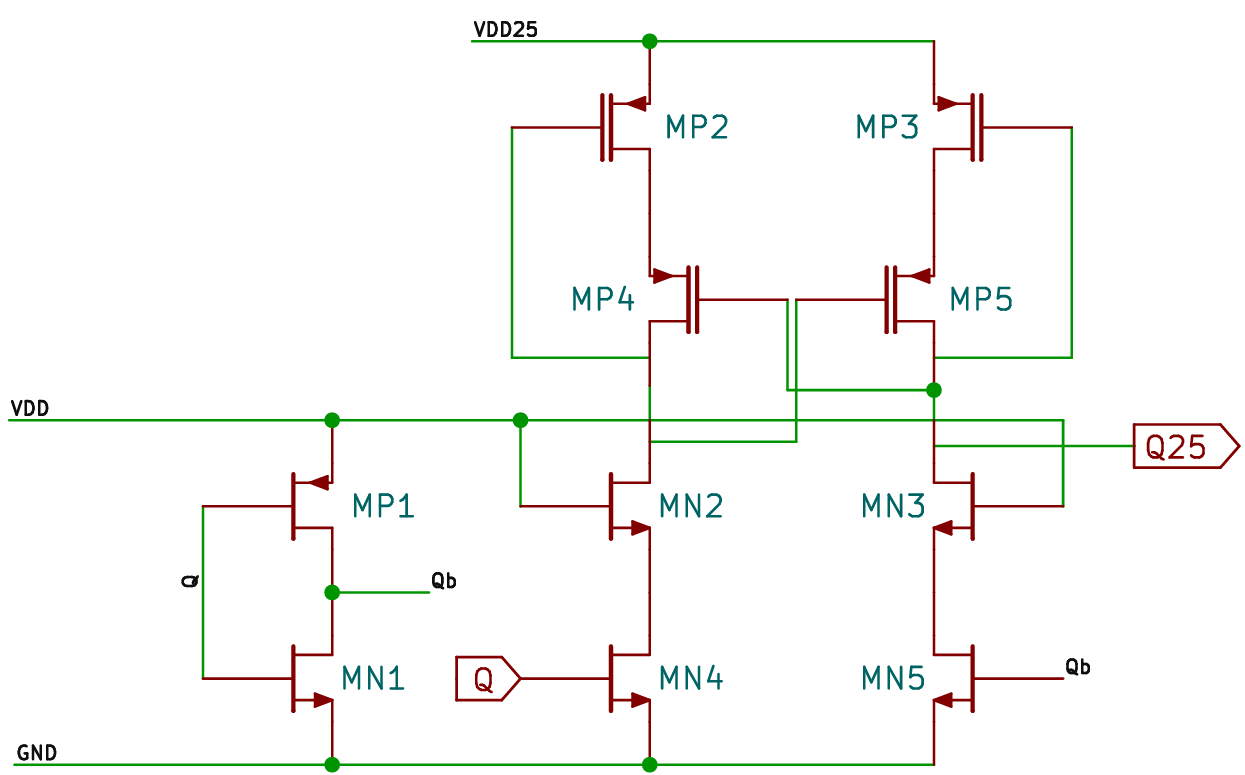}
\caption{Schematic of the voltage level shifter. All I/O pMOS devices (MP2, MP3, MP4, MP5) are minimum size to reduce $t_{\text{off}}$ and power consumption.}\label{vls}
\end{figure}

To keep the sampling switch's on-resistance($R_{\text{on}}$) low, we want to keep its gate voltage high. A 2.5 V I/O nMOS is used as the switch, as the $R_{on}$ of the minimal length 2.5 V I/O device is significantly less than that of the minimal length 1.2 V core device of the same width, given that the gate voltage is 2.5 V and 1.2 V respectively. Also, the input voltage range can be restricted to the bottom half of the drain voltage range of 2.5 V I/O nMOS, obviating the use of a complementary switch is not required. 
Instead, this design requires a 1.2 V to 2.5 V voltage level shifter (VLS, Fig. \ref{vls}) for each of sampling switches. The limiting factor in voltage uncertainty is the on-to-off transition time of the sampling switch $(t_{\text{off}})$; the transition time from off to on $(t_{\text{on}})$ is not as important. We adjusted the device widths of the VLS to prioritize reducing the former and achieved 15ps (Best Case) - 22ps (Worst Case) for $t_{\text{off}}$.

%\begin{figure}[htb]
%\centering
%\includegraphics[width=\linewidth]{whyVLS.png}
%\caption{Voltage plot of a voltage level %shifter.}\label{whyVLS}
%\end{figure}
\FloatBarrier
\subsection{Shift Register}
\begin{figure}[htb]
\centering
\includegraphics[width=\linewidth]{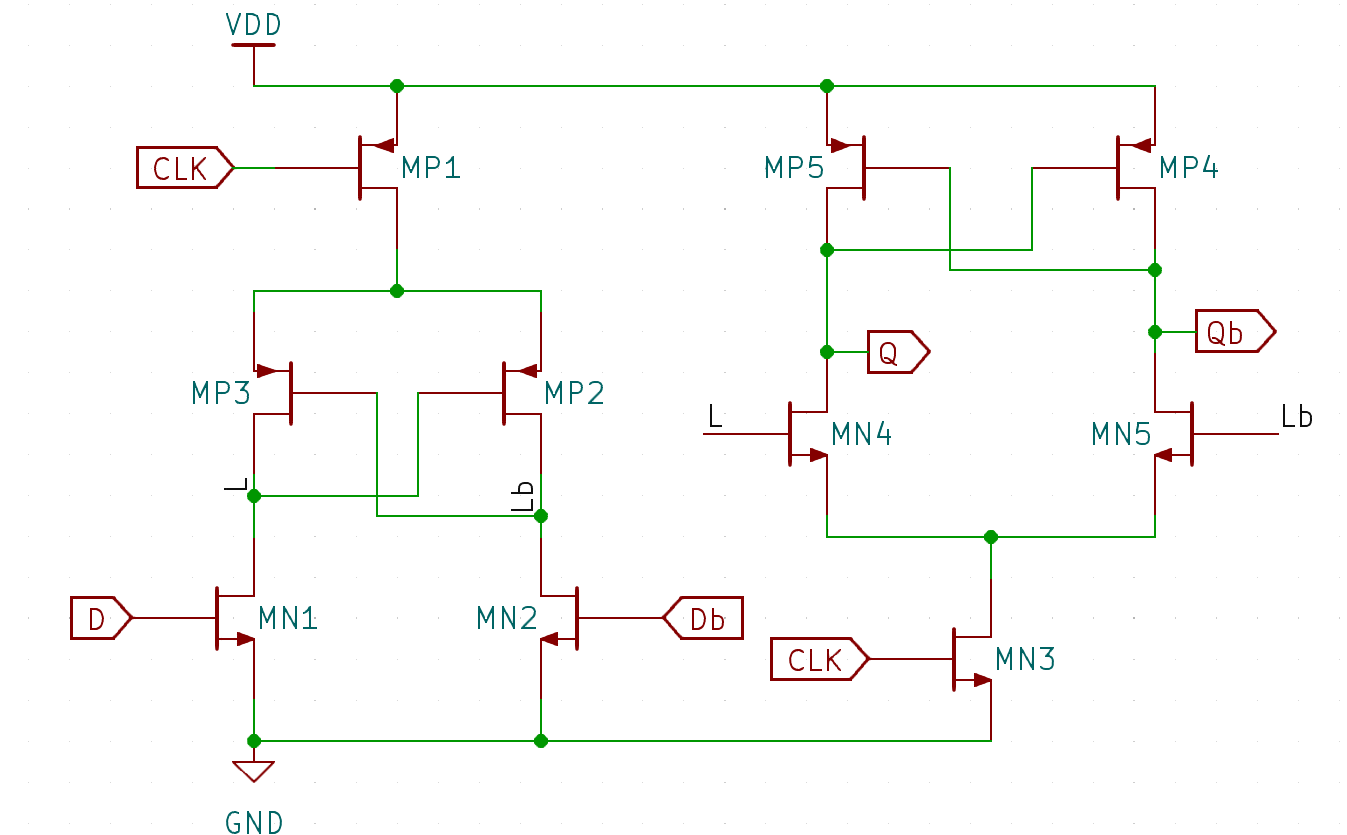}
\caption{Schematic of a dual edge-triggered filp flop.}\label{detff}
\end{figure}

To achieve a nominal 40 GS/s sampling rate of the interleaved SCA, we want each fast SCA column to operate at 10 GS/s. Delivering a 10 GHz clock to a wide area of the chip, however, results in high power consumption. Instead, we used an existing design \cite{diffFF} of a dual edge-triggered flip-flop (Fig. \ref{detff}) so that we only have to deliver a 5 GHz clock, halving the clock power consumption. 

\section{Signal Paths}
\label{signalSection}
\FloatBarrier

\subsection{Chip Entry}
Wire bonds and electrostatic protection devices (ESD) are the limiting factors of the analog bandwidth. 

\begin{figure}[htb]
\centering
\includegraphics[width=0.8\linewidth]{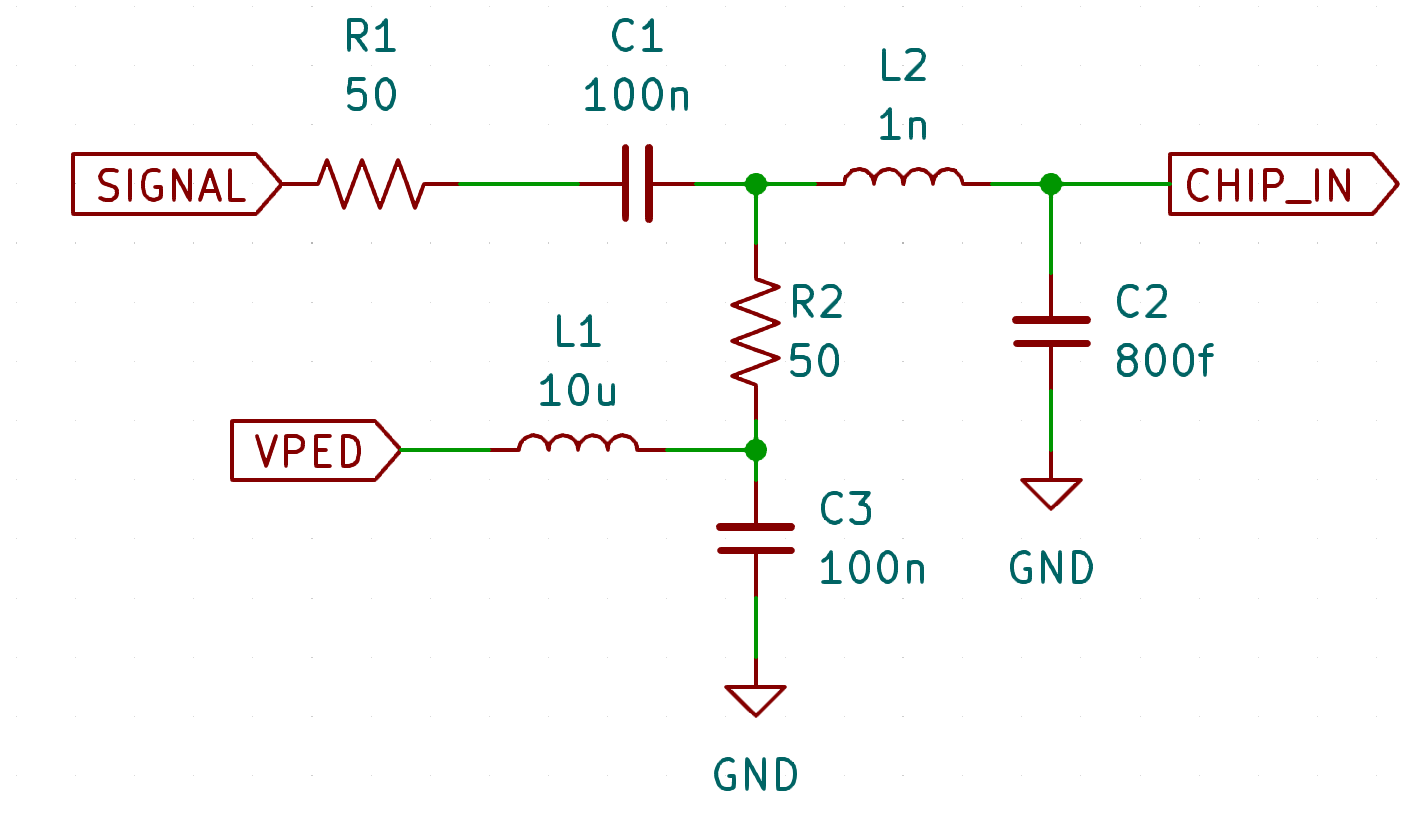}
\caption{Equivalent Schematic of the wire bond and ESD. R1 represents 50 $\Omega$ input. C1, C3, R2, L1 are components placed on board for capacitive coupling. L2 and C2 are the parasitic inductance and the capacitance of the wire bond and ESD combined. }\label{chipEntry}
\end{figure}

We employ a capacitively-coupled input to control the DC bias of the signal input (Fig. \ref{chipEntry}), which affects the analog bandwidth of the sampling switches. The lower the DC bias, the higher the analog bandwidth. Both the input and output source followers have limited voltage ranges, however, so the DC bias is set to approximately 600 mV.
\FloatBarrier
\FloatBarrier
\subsection{Main Signal Path}
\FloatBarrier

\begin{figure}[htb]
\centering
\includegraphics[width=0.4\linewidth]{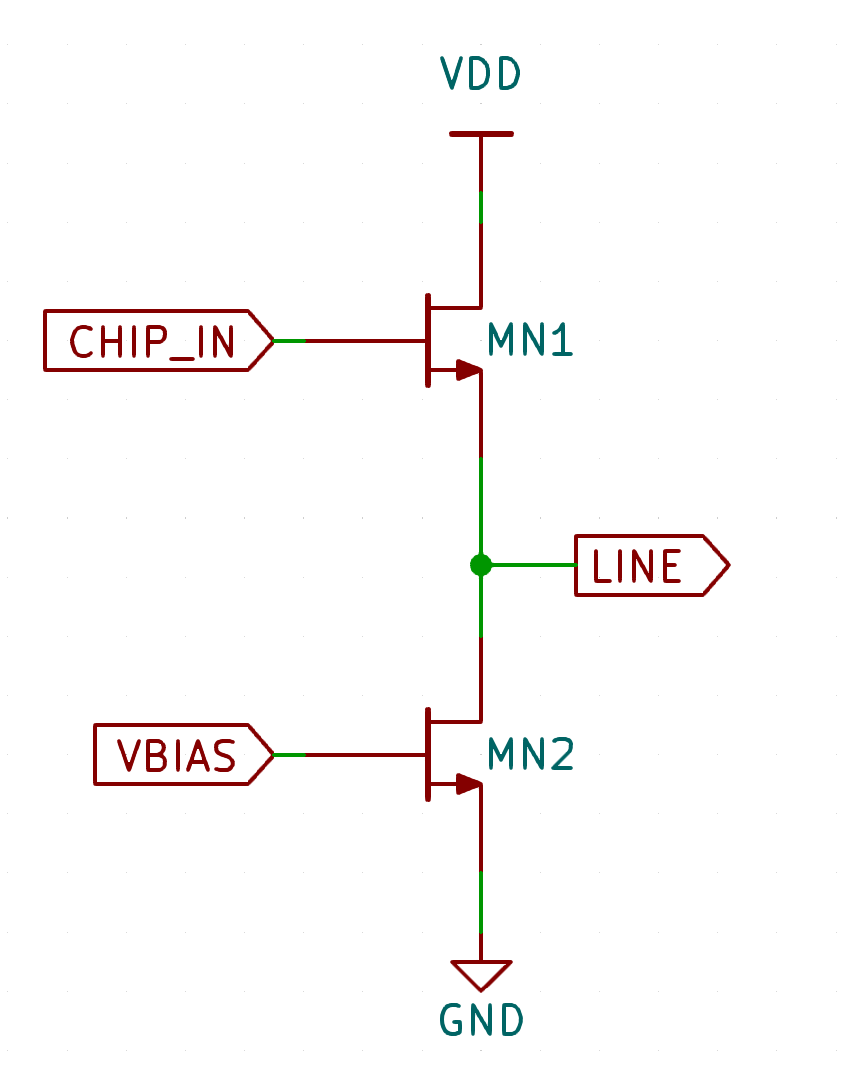}
\caption{Input Source Follower.}\label{mainSignalPath}
\end{figure}

Because of the input inductance, delivering the signal directly to sampling capacitors results in a relatively low analog bandwidth($<$2 GHz). We used a single transistor source follower (Fig. \ref{mainSignalPath}) to increase this to 4 GHz. The fast SCA columns are interleaved and hence the switching noise from a column may corrupt the capacitor voltage of a different column; we use a dedicated source follower per fast SCA column for the noise isolation. The entire slow bank receives the signal from another source follower since it is sampling at 5 GS/s and does not require a 4 GHz bandwidth.

\subsection{Readout}
\FloatBarrier

Since the sampling MOM capacitors are small (35fF), charge leakage before readout is an issue. We placed two stages of source follower that deliver the signal from the capacitor to the chip output without charge leakage.
%\begin{figure}[htb]
%\centering
%\includegraphics[width=0.5\linewidth]{Output_Follower_Picture.png}
%\caption{Output Source Follower.}\label{Readout}
%\end{figure}

\section{Physical Layout}
\label{layoutSection}
\FloatBarrier
The design has been laid out in a 2.4 mm $\times$ 1 mm rectangle, and the post-layout simulation to measure the performance is ongoing.

\begin{figure}[htb]
\centering
\includegraphics[width=0.7\linewidth]{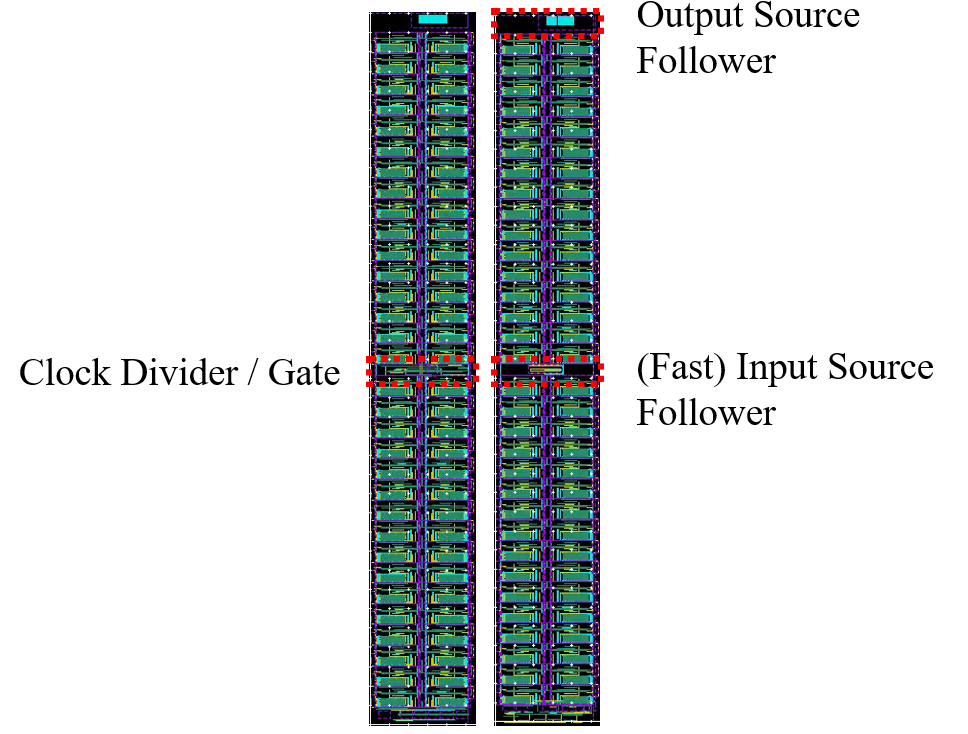}
\caption{Layout of a slow (Left) and fast (Right) SCA column.}\label{fastSCALayout}
\end{figure}

Both fast and slow SCA columns are laid out as 25.5$\mu$m$\times$200$\mu$m blocks (Fig. \ref{fastSCALayout}) with a dedicated clock source, where each column consists of 64 samples. There are 16 slow SCA columns and 4 fast SCA columns per channel, stacked horizontally. Naturally, slow SCA columns take four times the area of the fast SCA columns.

\FloatBarrier
\subsection{Clock Gating}
\begin{figure}[htb]
\centering
\includegraphics[width=\linewidth]{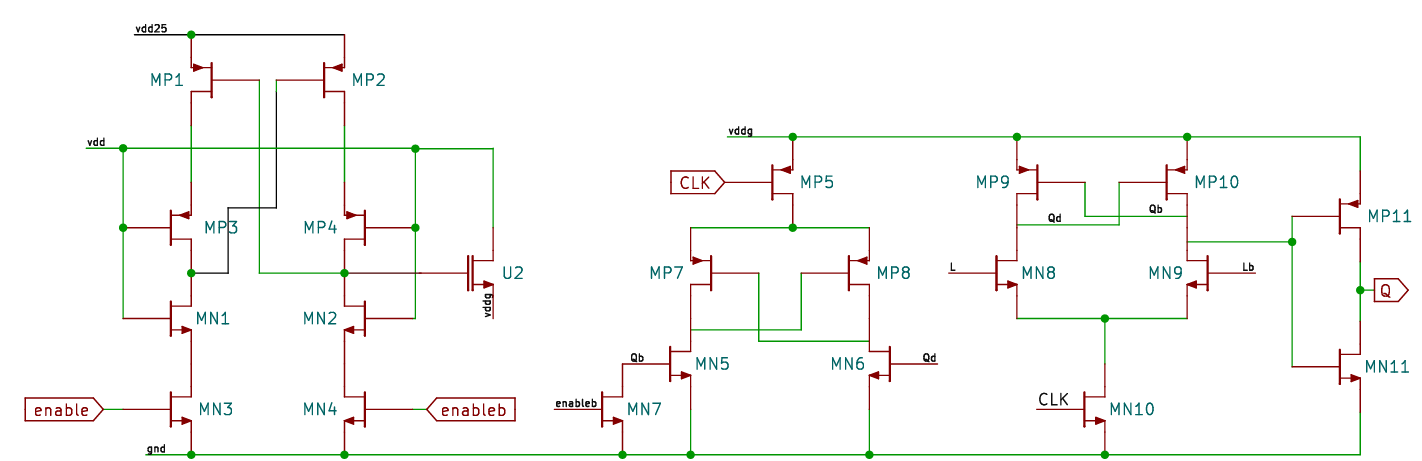}
\includegraphics[width=\linewidth]{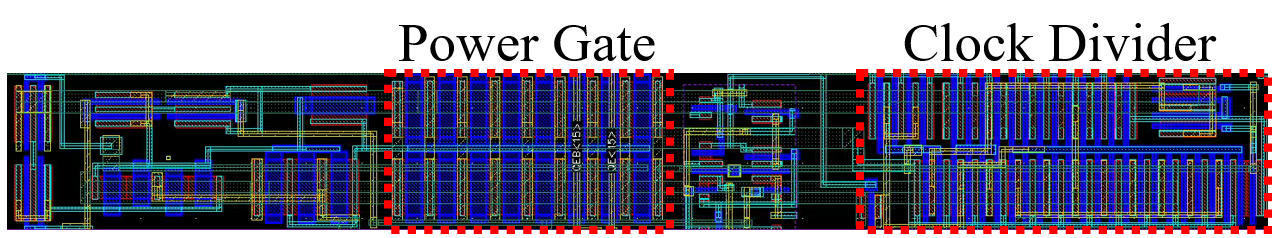}
\caption{Schematic (Top) and Layout (Bottom) of the clock gate/divider.}\label{clkdiv2Scheme}
\end{figure}

Unlike fast SCA columns, only one slow SCA column is active at any given time during sampling. To reduce clock power consumption, we designed a clock gate (Fig. \ref{clkdiv2Scheme}) so that a maximum of two slow SCA columns get the clock at any given time. This also doubles as a clock divider, as the slow columns operate at half the frequency of the fast columns.

\FloatBarrier
\section{Simulation Results}
\label{simSection}
\subsection{Input Voltage range}

\begin{figure}[htb]
\centering
\includegraphics[width=\linewidth]{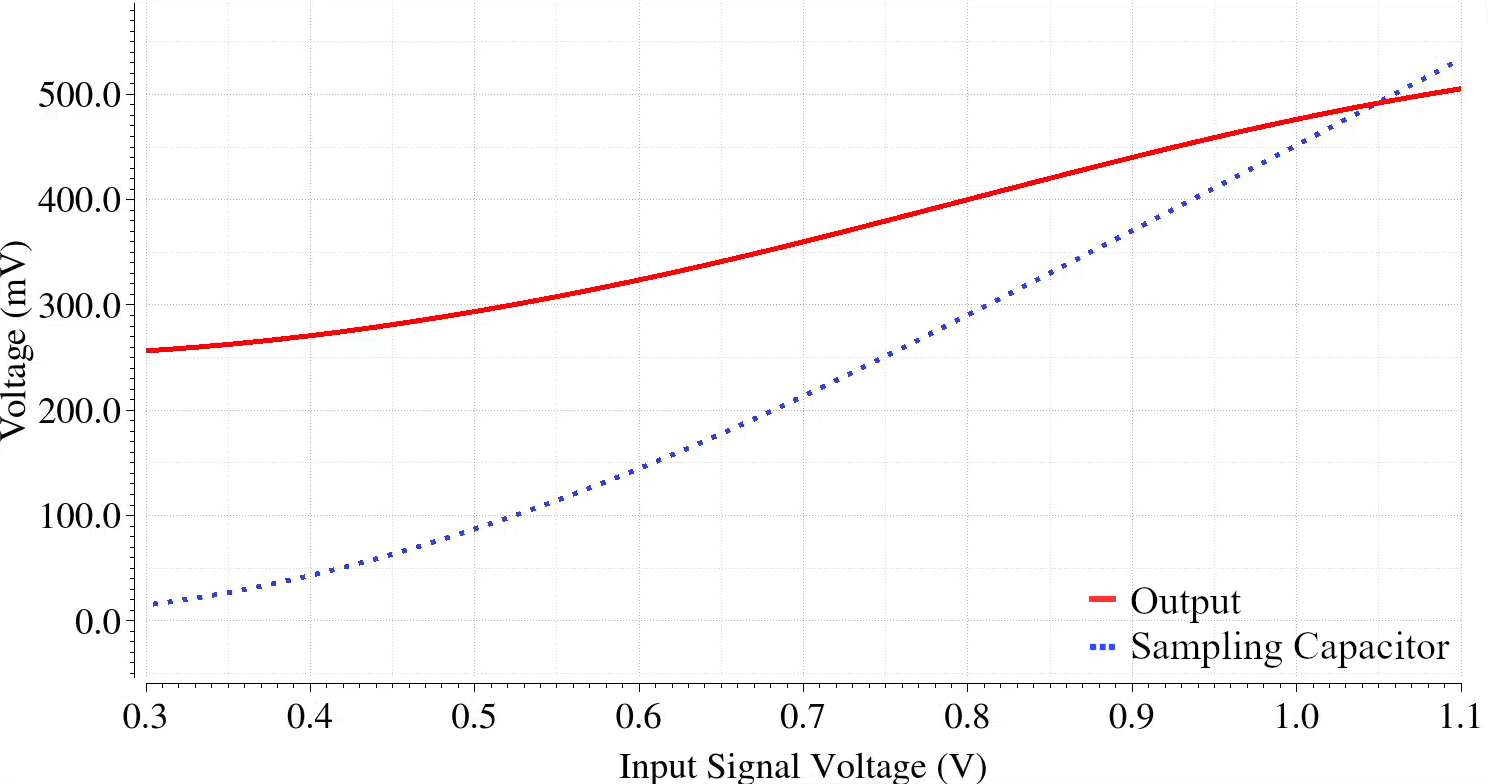}
\caption{Voltage plot of the source follower chain.}\label{sourceFollowerVoltagePlot}
\end{figure}

The simulation of the complete signal path, consisting of the input source follower, the capacitor, and the output source follower is shown in Fig. \ref{sourceFollowerVoltagePlot}. The rough linearity of the output is maintained for the input voltage range of 300 mV to 1.1 V. Calibration software may be used to record the curve and reconstruct the input voltage from the output.
\FloatBarrier
\subsection{Power consumption}
\FloatBarrier
The average power consumption in various process corners is simulated and shown in Table \ref{power}. The input source follower is turned off during the readout, further reducing the power consumption below 1 mW/Ch.
\begin{table}[htb]%% placement specifier
%% Use tabular environment to tag the tabular data.
%% https://en.wikibooks.org/wiki/LaTeX/Tables#The_tabular_environment
\centering%% For centre alignment of tabular.
\begin{tabular}{|c|c|c|}%% Table column specifiers
%% Tabular cells are separated by &
\hline
   & Worst Case & Best Case \\ %% A tabular row ends with \\
   \hline
   Input Source Follower [mW/Ch] & 9.2 & 4.0\\ %% A tabular row ends with \\

\hline
  SCA (Sampling) [mW/Ch] & 16.6 & 13.9 \\
\hline
\end{tabular}
%% Use \caption command for table caption and label.
\caption{Power consumption.}\label{power}
\end{table}
\section{Conclusion and Current Status}

All of the blocks in Figure $\ref{psec5Overview}$ have been laid out. Currently, we are running post-layout simulations to estimate the voltage and time uncertainty. The intended submission is before the end of 2024.

%% The Appendices part is started with the command \appendix;

%appendix sections are then done as normal sections
%\appendix
%\section{Example Appendix Section}
%\label{app1}

%Appendix text.

%Example citation, See \cite{friendly_ref_name}.
%%
  \bibliographystyle{elsarticle-num-names} 
  \bibliography{main}

\end{document}